**Concept of "post-eutectic" composition for undercooled alloys**

Andriy Gusak[(1,2,*)], Serhii Abakumov[(2)]

1- Centre of Excellence ENSEMBLE3 Sp. z o.o., ul. Wolczynska 133, 01-919 Warsaw, Poland
2- Cherkasy National University, Shevchenko blvd. 81, 18000, Cherkasy, Ukraine
    *- corresponding author

*Abstract*

*The concept of post-eutectic composition at a fixed undercooling is introduced as the most probable composition in zones of coupled growth. This composition corresponds to the minimum Gibbs free energy release rate in the steady-state regime, aligning with Prigogine's principle of minimum entropy production for steady states. Geometrically, this is represented by the "parallel tangents rule," a generalization of the "common tangent rule" for transitioning from equilibrium to steady-state conditions.*

The eutectic composition and the related eutectic temperature in binary alloys A-B are well-defined thermodynamic concepts corresponding to a three-phase equilibrium, where the composition of the liquid lies between the compositions of two solid phases. As described in textbooks, alloys with atomic fractions of B that are less than or greater than the eutectic composition are called hypoeutectic and hypereutectic, respectively. These alloys exhibit primary crystallization of the α phase (in the hypoeutectic case) or the β phase (in the hypereutectic case) during very slow cooling, before reaching the strictly eutectic composition at the strictly eutectic temperature. [1].

Under realistic conditions of finite (non-zero) cooling rates, crystallization occurs under non-equilibrium conditions with finite undercooling, $\Delta T$. We will refer to this situation as "**post-eutectic**." A natural question arises: can we introduce analogs of eutectic, hypoeutectic, and hypereutectic compositions under finite undercooling? At a fixed non-zero undercooling, $\Delta T$, the **eutectic** composition **point** $C_{eut}$ (the atomic fraction of component B in the liquid phase, corresponding to the common tangent for the three Gibbs energy-composition curves $g_\alpha(C, T_{eut}), g_{liq}(C, T_{eut}), g_\beta(C, T_{eut})$) transforms into a well-known composition **interval** $(C_{l\beta}, C_{l\alpha})$, with the margins determined by the equilibria between liquid-$\beta$ and $\alpha$-liquid. (Fig.1):

$$\frac{g_\beta - g_{liq}(C_{l\beta})}{C_\beta - C_{l\beta}} = \frac{\partial g_{liq}(C_{l\beta})}{\partial C_{liq}}, \frac{g_{liq}(C_{l\alpha}) - g_\alpha}{C_{l\alpha} - C_\alpha} = \frac{\partial g_{liq}(C_{l\alpha})}{\partial C_{liq}} \quad (1)$$

The question is – what point within this interval may be treated as the "**descendant**" of eutectic composition?

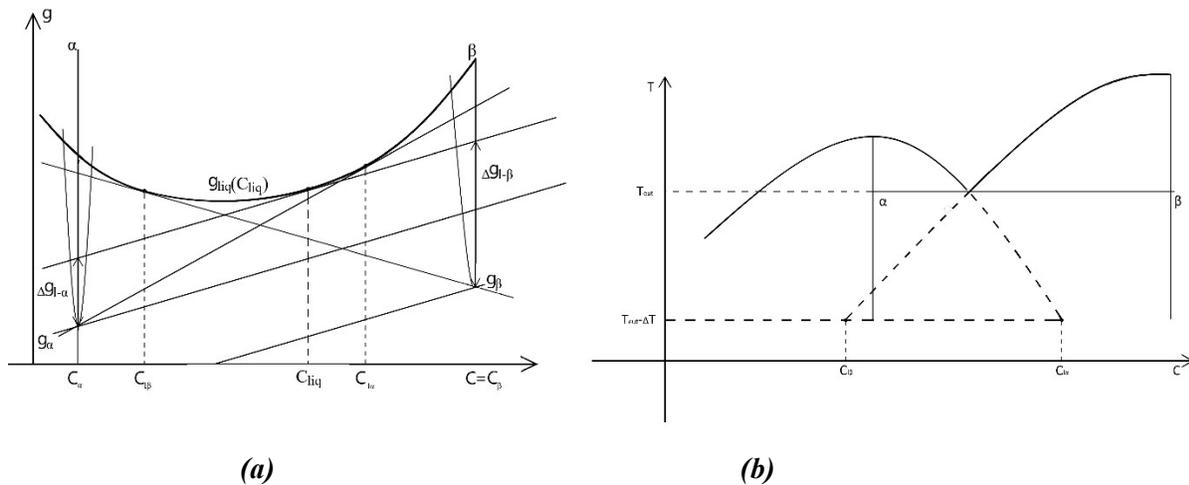

*(a)* *(b)*

*Fig.1. Post-eutectic interval. (a) Common tangents construction for finding the post-eutectic composition interval $(C_{l\beta}, C_{l\alpha})$. Here $C_{l\alpha}$, $C_{l\beta}$ are determined as the solutions of the equations $\frac{g_{liq}(C_{l\alpha})-g_{\alpha}}{C_{l\alpha}-C_{\alpha}} = \frac{\partial g_{liq}(C_{l\alpha})}{\partial C_{liq}}$, $\frac{g_{\beta}-g_{liq}(C_{l\beta})}{C_{\beta}-C_{l\beta}} = \frac{\partial g_{liq}(C_{l\beta})}{\partial C_{liq}}$. (b) Dependence of post-eutectic interval on undercooling.*

Let us consider the steady-state process under a fixed undercooling, $\Delta T$. For simplicity, we will treat the solid phases α and β as "point phases" (stoichiometric compounds or pure elements with practically zero solubility of the second element). A good example is the well-known InSb-Sb system [2], where the compound InSb represents the α-phase, and practically pure antimony represents the β-phase (Fig. 2).

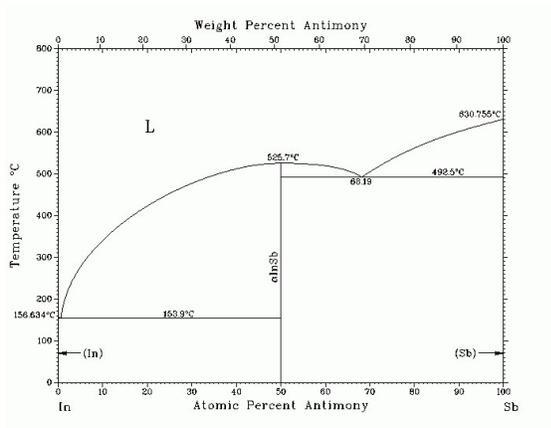

*Fig.2. Phase diagram of In-Sb system*

If one considers the plots of $C_{l\alpha}$ and $C_{l\beta}$ as functions of undercooling $\Delta T$, we obtain the post-eutectic part of the phase diagram, within which both phases **may** form simultaneously. The word "may" indicates that their formation is thermodynamically favorable, as it leads to a decrease in the Gibbs free energy of the alloy. In other words, both phases in this interval have the driving forces for nucleation and growth. **Can we predict the "most favorable" post-**

**eutectic composition within the post-eutectic interval?** – we try to answer this question in this Letter.

Based on our understanding of Prigogine's ideas on non-equilibrium thermodynamics for open systems [3,4], steady-state in an open system under non-uniform external conditions is analogous to the equilibrium state of a closed system under constant, uniform external conditions. For a closed system (a) isolated, or (b) in a thermal bath with fixed temperature T and volume V, or (c) in a thermal bath with fixed temperature T and pressure P, the system always tends (ignoring fluctuations) towards equilibrium. This corresponds to maximum entropy S in case (a), minimum Helmholtz free energy F=U-TS in case (b), and minimum Gibbs free energy G=U-TS+PV in case (c). In the case of an open system under constant but non-uniform boundary conditions (such as a constant temperature difference, chemical potential difference, or constant divergence of heat or material fluxes), the system may reach a steady state or exhibit oscillatory behavior. Examples of oscillatory behaviors, which are well-known in synergetic, include certain regimes of the predator-prey model, the Brusselator, the Lorenz model, and some oscillatory exothermic reactions. Here, we will focus on the **steady-state** as the final state of an open system. In some simple cases, the steady state corresponds to the minimum of entropy production (Prigogine principle). For an isolated closed system, the steady-state is reduced to equilibrium, and this minimum is just zero, and the entropy itself is maximal. It is important to note that for a system in a thermal bath with fixed temperature and pressure, the minimum entropy production for the unified total system (system + thermal bath) is equivalent to the minimum release rate of the Gibbs free energy (-1/T*dG/dt) of the system. For clarity, to avoid confusion, we provide proof of this statement in Appendix A, without claiming originality.

We will now see that the minimum release rate criterion can select a reasonable candidate for the post-eutectic composition after reaching a quasi-steady-state regime (a detailed proof is provided in Appendix B). Let $C_{liq}$ be the atomic fraction of the B component in the liquid alloy within the post-eutectic interval ($C_{l\beta}$, $C_{l\alpha}$). Let dN be the number of crystallized atoms. In Appendix B we show that the Gibbs free energy release per one crystallized atom can be found as

$$-\frac{dG}{dN} = \frac{C_\beta - C_{liq}}{C_\beta - C_\alpha}[g_{liq}(C_{liq}) - g_\alpha - (C_{liq} - C_\alpha)\frac{\partial g_{liq}(C_{liq})}{\partial (C_{liq})}] +$$

$$+ \frac{C_{liq} - C_\alpha}{C_\beta - C_\alpha}[g_{liq}(C_{liq}) - g_\beta + (C_\beta - C_{liq})\frac{\partial g_{liq}(C_{liq})}{\partial (C_{liq})}] \qquad (2)$$

Here, we treat this release rate as a function of $C_{liq}$. This function reaches its minimum when the first derivative is zero, and the second derivative is positive. In Appendix B, we demonstrate that this minimum condition can be formulated in two equivalent forms:

First form can be called "Parallel tangents rule" (Fig.3):

$$\frac{g_\beta - g_\alpha}{C_\beta - C_\alpha} = \frac{\partial g_{liq}\left(C_{liq}^{post-eut}\right)}{\partial C_{liq}} \qquad (3)$$

Second form of the minimum G-release rate can be called "equal driving forces per atom for two phases" (also see Fig.3): $\quad \Delta g_\beta = \Delta g_\alpha$, or $\qquad (4)$

$$g_{liq}(C^{post-eut}) - g_\alpha - (C^{post-eut} - C_\alpha)\frac{\partial g_{liq}(C^{post-eut})}{\partial(C_{liq})} =$$

$$= g_{liq}(C^{post-eut}) - g_\beta + (C_\beta - C^{post-eut})\frac{\partial g_{liq}(C^{post-eut})}{\partial(C_{liq})}$$

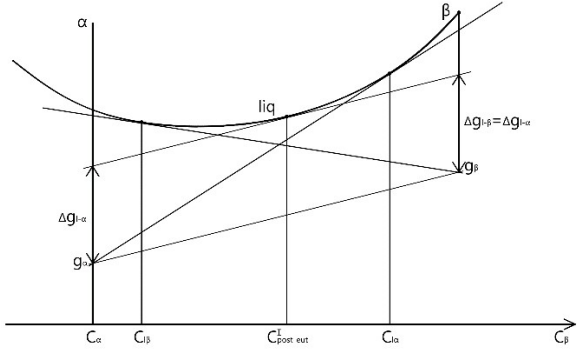

*Fig.3. "Parallel tangents rule" corresponding to minimum Gibbs free energy release rate and to equal driving forces*

.

The analytic approximation of the post-eutectic composition shift, expressed as a Taylor series expansion, is detailed in Appendix C:

$$\Delta C = \Delta T \frac{\frac{S_\beta - S_\alpha}{C_\beta - C_\alpha} - \frac{\partial S_{liq}}{\partial C}}{\frac{\partial^2 g_{liq}(T_{eut}, C_{eut})}{\partial C^2}} \qquad (5)$$

Notably, the shift in post-eutectic composition due to undercooling may be more pronounced in systems where there is a significant difference between the entropies of the crystalline phases.

Using the thermodynamic data for the In-Sb system [5] (Appendix D), the dependence of the post-eutectic composition difference, $C^{post-eut}(\Delta T) - C_{eut}$, on undercooling can be easily determined from eq. (3- 4) - Fig. 4a.

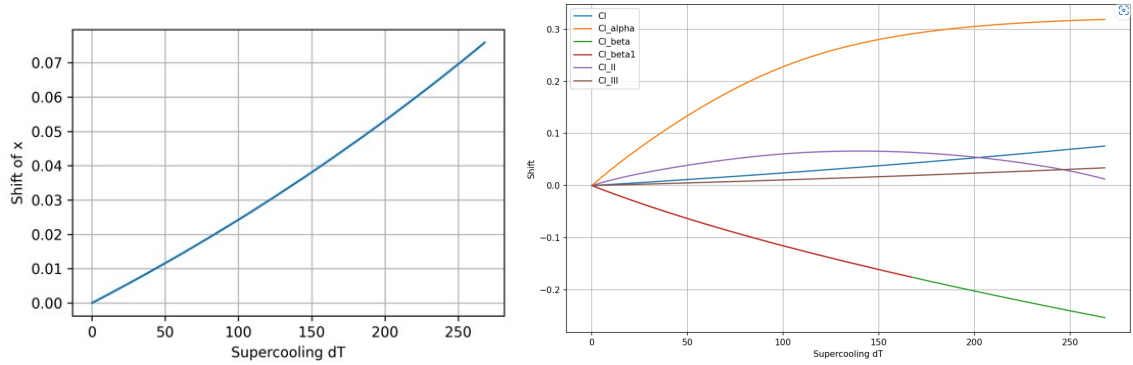

*Fig.4. Post-eutectic effects. (a) Dependence of difference $\Delta x = C_{post-eut} - C_{eut}$ on undercooling $\Delta T$ (C-atomic fraction of Sb). (b) Post-eutectic composition interval (in respect to Ceut) and three candidates for post-eutectic composition - appendixes 2(CI),5(CI-II),6-7 (CI-III, for the chosen parameters of surface tensions the plots obtained from appendixes 6 and 7, practically coincide)*

Note that in cases where one of the crystalline phases is a compound (as observed in the InSb-Sb system), significant undercooling can cause the post-eutectic interval to widen beyond the range defined by the eutectic couple, as depicted in Fig. 1b. This phenomenon implies that, under large enough undercooling, the crystallization process of a liquid alloy like InSb at equiatomic composition (which is "ready" to crystallize polymorphically, without requiring long-range diffusion) could also initiate the simultaneous nucleation of the Sb-phase.

Generally, any composition within the post-eutectic interval (illustrated in Fig. 1b) may result in the concurrent growth of a two-phase solidified alloy in forms such as colonies of lamellae or rods, among others. Conversely, if the nucleation barrier for one solid phase is significantly lower than that for the other (even at the eutectic composition of the liquid), dendrites or "plates" of the primary phase may initially form, followed by the coupled growth of "post-eutectic" colonies in the remaining liquid with a shifted "near-eutectic" composition.

We explored several other composition candidates favoring coupled growth using alternative criteria—such as equal growth rates (see Appendix E), equal nucleation barriers (see Appendix F), and equal nucleation rates (see Appendix G). Intuitively, our preference leans towards the first candidate, which corresponds to the minimum Gibbs free energy release rate (see Appendix B). Additional reason for this preference is the fact that the interlamellar or inter-rod distances within the coupled-growth colonies are usually much larger than the critical nucleation sizes, which means some "memory-loosing" effects, characteristic for general thermodynamic principles. [6]

In summary, the concept of post-eutectic composition, determined by the "parallel tangents rule" and associated with the minimum Gibbs free energy release rate under steady-state conditions, appears to be a reasonable approach for predicting the composition shift within coupled growth zones as a function of undercooling.

**ACKNOWLEDGMENTS**

We are grateful for the "ENSEMBLE3-Center of Excellence for nanophononics, advanced materials and novel crystal growth-based technologies" project (GA No. MAB/2020/14) carried out under the International Research Agenda programs of the Foundation for Polish Science that are co-financed by the European Union under the European Regional Development Fund and the European Union Horizon 2020 research and innovation program Teaming for Excellence (GA. No. 857543) for supporting this work. The publication was created as part of the project of the Minister of Science and Higher Education "Support for the activities of Centers of Excellence established in Poland under the Horizon 2020 program" under contract No.MEiN/2023/DIR/3797.

**Appendix A. Interrelation between entropy production and Gibbs free energy release rate.**

Let us consider a thermodynamic system with energy E and volume V and fixed number of atoms of each component in the thermal bath (with much larger energy and volume) characterized by temperature T and pressure P, so that in total (system and thermal bath) the united system is isolated with fixed energy $E_{total}$, and volume $V_{total}$. If we neglect the input of interface, then $E_{bath} = E_{total} - E$, $V_{bath} = V_{total} - V$ (it is important that $E_{total} \gg E, V_{total} \gg V$)

$$S_{tot} = S(E_{tot} - E, V_{tot} - V) + S(E, V) =$$

$$= S(E_{tot}, V_{tot}) + S(E, V) - E\frac{\partial S}{\partial E} - V\frac{\partial S}{\partial V} = S(E_{tot}, V_{tot}) + S(E, V) - \frac{E}{T} - V\frac{P}{T} = S(E_{tot}, V_{tot}) - \frac{E+PV-S}{T} = const - \frac{G}{T}. \quad (A1)$$

**Thus, $\frac{dS_{total}}{dt} = -\frac{1}{T}\frac{dG}{dt}$ – at fixed temperature and pressure
the entropy production in the whole united system is equal to the release rate of Gibbs free energy in the given system.**

**Appendix B. Application of the Prigogine's principle of the "Minimum entropy production" – "Rule of parallel tangents" (instead of common tangents)**

Let us analyze the decrease rate of Gibbs free energy due to decomposition of binary liquid with atomic fraction arbitrary but fixed Cliq of the component B into crystalline compounds alpha and beta. For this let us at first consider the Gibbs free energy change due to growth of single phase "alpha" under constraint of fixed total number of the atoms B in the system:

$$dG = d(N_\alpha g_\alpha + (N^{total} - N_\alpha)g_{liq}(C_{liq})) \quad , \quad (B1)$$

$$0 = d(N_\alpha C_\alpha + (N^{total} - N_\alpha)C_{liq}) = dN_\alpha(C_\alpha - C_{liq}) + (N^{total} - N_\alpha)dC_{liq} =>$$

$$=> (N^{total} - N_\alpha)dC_{liq} = -(C_\alpha - C_{liq})dN_\alpha \quad (B2)$$

$$dG = d\left(N_\alpha g_\alpha + (N^{total} - N_\alpha)g_{liq}(C_{liq})\right) = dN_\alpha(g_\alpha - g_{liq}(C_{liq})) + \frac{\partial g_{liq}(C_{liq})}{\partial(C_{liq})}(N^{total} - N_\alpha)dC_{liq} = -\Delta g_\alpha dN_\alpha \quad (B3)$$

Here $\Delta g_\alpha = g_{liq}(C_{liq}) - g_\alpha - (C_{liq} - C_\alpha)\frac{\partial g_{liq}(C_{liq})}{\partial(C_{liq})}$.  (B4)

Similarly, growth of beta-compound will give the decrease of Gibbs free energy $dG = -\Delta g_\beta dN_\beta$

with $\Delta g_\beta = g_{liq}(C_{liq}) - g_\beta + (C_\beta - C_{liq})\frac{\partial g_{liq}(C_{liq})}{\partial(C_{liq})}$  (B5)

Two driving forces per atom of new phase (B4) and (B5) are shown at the Figure 1a.

If two compounds are growing simultaneously transforming dN atoms of liquid into $dN_\alpha$ atoms of alpha-phase plus $dN_\beta$ atoms of beta-phase. Then thermodynamics plus conservation of matter give:

$dG = -\Delta g_\alpha dN_\alpha - \Delta g_\beta dN_\beta$,

$(dN_\alpha + dN_\beta)C_{liq} = C_\alpha dN_\alpha + C_\beta dN_\beta \Rightarrow (C_{liq} - C_\alpha)dN_\alpha = (C_\beta - C_{liq})dN_\beta \Rightarrow$

$\Rightarrow dN_\alpha = \frac{C_\beta - C_{liq}}{C_\beta - C_\alpha}dN, \quad dN_\beta = \frac{C_{liq} - C_\alpha}{C_\beta - C_\alpha}dN$ (lever rule),  (B6)

$$-\frac{dG}{dN} = \frac{C_\beta - C_{liq}}{C_\beta - C_\alpha}[g_{liq}(C_{liq}) - g_\alpha - (C_{liq} - C_\alpha)\frac{\partial g_{liq}(C_{liq})}{\partial(C_{liq})}] +$$

$+ \frac{C_{liq} - C_\alpha}{C_\beta - C_\alpha}[g_{liq}(C_{liq}) - g_\beta + (C_\beta - C_{liq})\frac{\partial g_{liq}(C_{liq})}{\partial(C_{liq})}]$.  (B7)

Let us remind that the $(-\frac{dG}{dN})$ in eq. (B7) is a release of Gibbs free energy due per one atom (or one mole) during crystallization of undercooled liquid at the temperature $T = T_{eut} - \Delta T$. It is simultaneously the entropy production in the Universe due to this crystallization. It seems natural (bearing in mind the Prigogine's principle of minimum entropy production for the steady-state open systems) to check the extremum properties of $(-\frac{dG}{dN})$. It is easy to check that

$\frac{\partial}{\partial C_{liq}}\left(-\frac{dG}{dN}\right) = \frac{\Delta g_\beta - \Delta g_\alpha}{C_\beta - C_\alpha}$ with driving forces determined by eq. (A4, A5).  (B8)

$\frac{\partial^2}{\partial C_{liq}^2}\left(-\frac{dG}{dN}\right) = \frac{\partial^2 g_{liq}}{\partial C_{liq}^2} > 0$ for metastable liquid.  (B9)

Thus, the condition of equal driving forces per atom for both crystalline phases,

$\Delta g_\beta = \Delta g_\alpha$,  (B10)

corresponds to the minimum entropy production in respect to variable composition of the liquid phase. Condition (B10), after substitution of eqs (B4, B5), can be formulated as a "parallel tangents rule":

This is our first and main candidate for the role of "post-eutectic" as a composition "favorable" for coupled growth at the fixed undercooling.

Note that the rule of parallel tangents in the form (B10) has one more possible interpretation: bulk driving forces of nucleation of two compounds are equal – of course, this fact correlates with the simultaneous phase formation, if one neglects the surface energy effects.

# APPENDIX C. Analytic approximation of post-eutectic composition shift $\Delta C_{liq}^I = C_{liq}^I - C_{eut}$ with undercooling increase.

The main equation for this version of post-eutectic is:

$$\frac{\partial g_{liq}(T_{eut}-\Delta T, C^*=C_{eut}+\Delta C)}{\partial C} = \frac{g_\beta(T_{eut}-\Delta T) - g_\alpha(T_{eut}-\Delta T)}{C_\beta - C_\alpha} \tag{C1}$$

Let us approximate the dependence $g_{liq}(T_{eut} - \Delta T, C^* = C_{eut} + \Delta C)$ by the Taylor expansion up to second-order terms in respect to composition deviation from eutectic one, and up to first-order terms in respect to undercooling. Also, we regard the phases alpha and beta as "point-like" in respect to composition and expand them up to first-order terms in respect to temperature:

$$\frac{\partial g_{liq}(T_{eut},C_{eut})}{\partial C} - \Delta T \frac{\partial^2 g_{liq}(T_{eut},C_{eut})}{\partial C \partial T} + \Delta C \frac{\partial^2 g_{liq}(T_{eut},C_{eut})}{\partial C^2} = \frac{g_\beta(T_{eut}) - \Delta T \frac{\partial g_\beta(T_{eut})}{\partial T} - g_\alpha(T_{eut}) + \Delta T \frac{\partial g_\alpha(T_{eut})}{\partial T}}{C_\beta - C_\alpha} =$$

$$\frac{g_\beta(T_{eut}) - g_\alpha(T_{eut})}{C_\beta - C_\alpha} + \Delta T \frac{S_\beta - S_\alpha}{C_\beta - C_\alpha} \tag{C2}$$

The first terms in the left- and right-hand sides correspond to equilibrium condition for three phases alpha, beta and liquid at eutectic temperature $T_{eut}$, so that they are cancelled. The rest of this equation gives:

$$\Delta T \frac{\partial S_{liq}}{\partial C} + \Delta C \frac{\partial^2 g_{liq}(T_{eut},C_{eut})}{\partial C^2} = \Delta T \frac{S_\beta - S_\alpha}{C_\beta - C_\alpha}, \tag{C3}$$

so that $\quad \Delta C = \Delta T \dfrac{\frac{S_\beta - S_\alpha}{C_\beta - C_\alpha} - \frac{\partial S_{liq}}{\partial C}}{\frac{\partial^2 g_{liq}(T_{eut},C_{eut})}{\partial C^2}} \tag{C4}$

# Appendix D. Thermodynamic data for InSb-Sb system

According to [5] the molar Gibbs free energy dependence on composition and temperature for liquid phase can be approximated as

$g_{liq}$(T,C=Cat-sb) = C * (1 - C) * [(1 - C) * (a1 + a2 * T + a3 * T * ln(T)) +

C * (a4 + a5 * T +a6 * T * ln(T))+ C * (1 - C) * (a7 + a8 * T + a9 * C * T)]+

+ R * T * (x * ln(x) + (1 - x) * ln(1 - x))

Here a1 = -32623.14938503529,  a2 = 159.5916858754045,  a3 = -20.71718090599029, a4 = -22719.71923073586,  a5 = 121.1765817901414,  a6 = -15.53738056036172, a7 = -11638.31187890400,  a8 = -16.53375104522048,  a9 = 15.13297239974727 (Joule per mole of atoms)

For crystalline phases  InSb (α) and Sb (β):

$g_{In}$ =-(3263.52 - 7.5934 * T),  $g_\beta = g_{Sb}$ = -(19874 - 21.9868 * T),

$g_\alpha = g_{InSb} = g_{liq}$ (C=0.5, T) - (48097 - 60.215 * T) * 0.5

## Appendix E. Condition of constant composition in the remaining liquid during the simultaneous growth of two spherical phase particles.

$$(C_\alpha - C_{liq}^{II}) \frac{dR_\alpha^3}{\Omega_\alpha dt} + (C_\beta - C_{liq}^{II}) \frac{dR_\beta^3}{\Omega_\beta dt} = 0$$

$$\frac{dR_\alpha}{dt} = \frac{D_{liq}(C_{l\alpha} - C_{liq}^{II})}{(C_{l\alpha} - C_\alpha)R_\alpha}, \quad \frac{dR_\beta}{dt} = \frac{D_{liq}(C_{liq}^{II} - C_{l\beta})}{(C_\beta - C_{l\beta})R_\beta}.$$

Here $C_{l\alpha}, C_{l\beta}$ are determined by the numeric solution of the eqs (1).

The value of $C_{liq}^{II}$ is then determined by the numeric solution of the equation

$$\frac{(C_{liq}^{II} - C_\alpha)^2 (C_{l\alpha} - C_{liq}^{II})^3}{(C_{l\alpha} - C_\alpha)^3} = \frac{(C_\beta - C_{liq}^{II})^2 (C_{liq}^{II} - C_{l\beta})^3}{(C_\beta - C_{l\beta})^3} \frac{\Omega_\beta}{\Omega_\alpha}$$

Here $\Omega_\alpha = \Omega_{at}^{InSb} = 3.401 * 10^{-29} m^3/atom$, $\Omega_\beta = \Omega_{atom}^{Sb} = 3.024 * 10^{-29} m^3/atom$.

## Appendix F. Condition of equal nucleation barriers for homogeneous nucleation of both phases.

$$\frac{[g_{liq}(C_{liq}^{III}) - g_\alpha - (C_{liq}^{III} - C_\alpha) \frac{\partial g_{liq}(C_{liq}^{III})}{\partial C_{liq}}]^2}{\gamma_{\alpha l}^3 (\Omega_\alpha)^2} = \frac{[g_{liq}(C_{liq}^{III}) - g_\beta - (C_{liq}^{III} - C_\beta) \frac{\partial g_{liq}(C_{liq}^{III})}{\partial C_{liq}}]^2}{\gamma_{\beta l}^3 (\Omega_\beta)^2}.$$

For illustration, we treated the case $\gamma_{\alpha l} = \gamma_{\beta l}$.

## Appendix G. Condition of equal nucleation rates for both phases.

Applying the standard Zel'dovich theory for steady-state nucleation, we equalize the nucleation rates for precipitation in the supersaturated alloy [7, 8]

$$\frac{\gamma_{\alpha l}^3 (\Omega_\alpha)^2}{\left[g_{liq}(C_{liq}^{III}) - g_\alpha - (C_{liq}^{III} - C_\alpha) \frac{\partial g_{liq}(C_{liq}^{III})}{\partial C_{liq}}\right]^2} - \frac{\gamma_{\beta l}^3 (\Omega_\beta)^2}{\left[g_{liq}(C_{liq}^{III}) - g_\beta - (C_{liq}^{III} - C_\beta) \frac{\partial g_{liq}(C_{liq}^{III})}{\partial C_{liq}}\right]^2} = \frac{3kT}{16\pi} \ln \left[ \sqrt{\frac{\gamma_{\alpha l}}{\gamma_{\beta l}}} \frac{C_\beta - C_{l\beta}}{C_{l\alpha} - C_\alpha} \right]$$